\documentclass[journal,dvips]{IEEEtran}

\usepackage[pdf]{pstricks}
\usepackage{pst-plot}
\usepackage{pstricks-add}
\usepackage{overpic}
\usepackage{pst-eps}

\usepackage[latin1]{inputenc}
\usepackage{graphicx} 
\usepackage{amsmath} 
\usepackage{amssymb}  
\usepackage{mathtools}
\mathtoolsset{showonlyrefs}
\usepackage{flushend}
\usepackage{url}
\usepackage{color}

\IEEEoverridecommandlockouts

\def\mat#1{{\mathbf #1}} 

\mathtoolsset{showonlyrefs}

\begin{document}

\author{Sebastian~Dingler$^1$ (orcid: 0000-0002-0162-8428) \thanks{$^1$This paper is an independent contribution without affiliation. To mail me remove foobar \tt\small s.dinglerfoobar@gmail.com}  }

\title{Fitting ellipses to noisy measurements}

\maketitle

\begin{abstract}
This work deals with fitting of ellipses to noisy measurements.
The literature knows many different approaches for this.
The main representatives are presented and discussed in this paper.
Furthermore, the case is considered when outliers are present in the measurement data.
Robust methods which are less sensitive to outliers are suitable for this case.
All discussed methods are compared by a simulation.
The code for the simulation is available for free use on github.com/sebdi/ellipse-fitting.
\end{abstract}


\section{Introduction}

\IEEEPARstart{A}{pproximating} given data using a model is a typical task in the context of pattern recognition, tracking and image processing.
The goal is to reduce the complexity of the data and to provide higher application layers with a simpler representation of the data.
Occasionally it is essential to estimate a simple model with measurements.
As an example the tracking of objects can be named.
In case of high measurement noise, few measurements or objects with a strong motion it is difficult to estimate the shape of an object.
In this case it is necessary to use a simple shape with few degrees of freedom to describe the object.
For this purpose the ellipse is one of the most used models.
In image processing ellipses often appear because a circle becomes an ellipse by perspective projection \cite{Porrill:1990,Frosio,Ahn}.
The same is true for spherical objects \cite{Greggio}.
Moreover, because of the non-linearity of the ellipse, it is a suitable model problem to test and compare different regression algorithms.
Due to the non-linearity, several methods exist to fit ellipses to measurements.
Each method uses a different assumption or simplification.
In this paper, these methods will be presented. 

To this end, Section \ref{sec:1} lays the necessary groundwork. Sections \ref{sec:lsfad} through \ref{sec:robust} then present the various methods. A simulation in Section \ref{sec:simulation} and a summary with outlook in Section \ref{sec:zus} complete this work.

\section{Mathematical description of ellipses}
\label{sec:1}
An ellipse $C$ can be described for the coordinate axes $x$ and $y$ as follows
\begin{equation}
C(x,y)=\mat{a} x^2 + 2 \mat{b} x y + \mat{c} y^2 + 2 \mat{d}x + 2\mat{e}y + \mat{f}=0\enspace .
\end{equation}
Here $\mat{a}$, $\mat{b}$ and $\mat{c}$ cannot be zero at the same time.
Another constraint is that $\mat{b}^2-\mat{a}\mat{c}<0$ must hold.
Given noisy measurements for this ellipse, the goal is to determine the parameters $\mat{a}$, $\mat{b}$, $\mat{c}$, $\mat{d}$, $\mat{e}$ and $\mat{f}$ in such a way that they best describe the data in terms of a quality criterion.
Because the trivial solution $\mat{a} = \mat{b} = \mat{c} = \mat{d} = \mat{e} = \mat{f} = 0$ must be prevented, the literature often uses three normalizations. 
These are:
\begin{itemize}
\item Normalization with $\mat{a}+\mat{c}=1$
\item Normalization with $\mat{a}^2+\mat{b}^2+\mat{c}^2+\mat{d}^2+\mat{e}^2+\mat{f}^2=1$
\item Normalization with $\mat{f}=1$.
\end{itemize}
Another useful form of the elliptic equation is the general parametric form
\begin{align} 
x(t) &= x_c + a \ \text{cos}(t) \ \text{cos}(\alpha)- b \ \text{sin}(t) \ \text{sin}(\alpha) \enspace ,  \label{eq:paraForm1} \\
y(t) &= y_c + a \ \text{cos}(t) \ \text{sin}(\alpha)+ b \ \text{sin}(t) \ \text{cos}(\alpha) \enspace , \label{eq:paraForm2}
\end{align}
where the parameter $t \in [0,2 \pi]$ can be varied to get different values of the ellipse. 
The other parameters are the center of the ellipse with $x_c$ and $y_c$, the semi-major axis $a$, the semi-minor axis $b$ and $\alpha$ the angle between $x$-axis and the semi-major axis $a$. 

Furthermore, an ellipse can also be represented without rotation and a center at the origin of the coordinate system.
This coordinate system is denoted with $XY$ to distinguish it from the general coordinate system $xy$.
The transformation between the coordinate systems $xy$ and $XY$ is calculated with the rotation matrix
\begin{eqnarray}
\mat{R}=\begin{bmatrix}
       \text{cos} \alpha & \text{sin} \alpha \\[0.3em]
       - \text{sin} \alpha & \text{cos} \alpha
     \end{bmatrix} \enspace .
\end{eqnarray}
The transformation for a point $\mat{x}=[x,y]^T$ in the $xy$ coordinate system is then done with
\begin{eqnarray}
\mat{X}=\mat{R}(\mat{x}-\mat{x}_c) \enspace ,
\end{eqnarray}
with $\mat{x}_c=[x_c,y_c]^T$ and $\mat{X}=[X,Y]^T$.
The reverse transformation with
\begin{eqnarray}
\mat{x}=\mat{R}^{-1}\mat{X}+\mat{x}_c \enspace .
\end{eqnarray}

\section{Least-squares fitting based on algebraic distance}
\label{sec:lsfad}
In this section, the least-squares method will be used to show how the parameters of the ellipse can be determined based on noisy measurements.
Applying the normalization $\mat{a}+\mat{c}=1$ to the elliptic equation, this eliminates $\mat{c}$ and leads to
\begin{eqnarray}
C(x,y) = \mat{a} (x^2 - y^2) + 2 \mat{b} x y  + 2 \mat{d}x + 2\mat{e}y + y^2 + \mat{f} \enspace .
\end{eqnarray}
This can be converted into vector notation with $\mat{\phi}=[x^2 - y^2, 2x y, 2 x, 2y, 1]$, $\mat{p}=[\mat{a}, \mat{b}, \mat{d}, \mat{e}, \mat{f}]^T$ and $z=-y^2$, to
\begin{eqnarray}
C(x,y) = \mat{\phi} \mat{p} -z .
\end{eqnarray}
If $n$ measurements $\left\{ (x_1,y_1), (x_2,y_2), ...., (x_n,y_n) \right\}$ are given, this leads to $n$ equations
\begin{eqnarray}
C_1(x_1,y_1) &=& \mat{\phi}_1 \mat{p} -z_1 \\
C_2(x_2,y_2) &=& \mat{\phi}_2 \mat{p} -z_2 \\
\vdots &=& \vdots \\
C_n(x_n,y_n) &=& \mat{\phi}_n \mat{p} -z_n \enspace .
\end{eqnarray}
These can be expressed in vector notation. For this purpose the following is defined 
\begin{eqnarray}
\mat{C} &=&[C_1(x_1,y_1),C_2(x_2,y_2),..,C_n(x,y)]^T \enspace ,\\
\mat{\Phi} &=&[\mat{\phi}_1^T,\mat{\phi}_2^T,..,\mat{\phi}_n^T]^T \enspace ,\\
\mat{y} &=& [z_1,z_2,...,z_n]^T \enspace .
\end{eqnarray}
This yields the equation
\begin{equation}
\mat{C}=\mat{\Phi} \mat{p} - \mat{y} \enspace .
\end{equation}
The goal is now to find a $\mat{p}$ for noisy measurements that optimally represents these measurements with respect to the least-squares criterion.
Since measurements are always noisy, there will usually be no $\mat{p}$ that is zero for all measurements $\mat{C}$, hence an error $\tilde{\mat{e}}$ with $\tilde{\mat{e}}=[\tilde{e}_1,\tilde{e}_2,...,\tilde{e}_n]^T$ will remain. Therefore, the real goal is to minimize the error $\tilde{\mat{e}}$ by choosing $\mat{p}$ appropriately. For this purpose, linear regression finds a vector $\mat{p}$ by minimizing the quality function
\begin{eqnarray}
\mat{G}(\mat{p}) &=& \mat{C}^T \mat{C} \\
&=& \tilde{\mat{e}}^T\tilde{\mat{e}} \\
&=& (\mat{\Phi} \mat{p} -\mat{y})^T (\mat{\Phi} \mat{p} -\mat{y}) \enspace .
\end{eqnarray}
This is achieved by setting the derivative of $\mat{G}$ equal to zero, resulting in
\begin{eqnarray}
0  &=& \frac{\partial}{\partial \mat{p}} \mat{G}(\mat{p}) \\
0  &=& \frac{\partial}{\partial \mat{p}} (\mat{\Phi} \mat{p} -\mat{y})^T (\mat{\Phi} \mat{p} -\mat{y}) \\
0 &=& 2 \mat{\Phi}^T (\mat{\Phi} \mat{p} - \mat{y})   \\
\mat{p} &=& (\mat{\Phi}^T \mat{\Phi})^{-1} \mat{\Phi}^T \mat{y} \enspace .
\end{eqnarray}
Thus $\mat{p}$ is the solution with the smallest squared error with respect to the algebraic distance since $C(x,y)$ is minimized. 
\section{Least-squares fitting based on orthogonal distance}
\begin{figure}[htbp]
    \centering
\begin{pspicture}(-4.5,-2.5)(3.5,3)
 
  \uput[45](0.6,2.2){$(x_i,y_i)$}
  \uput[45](0,0.1){$(x_c,y_c)$}
  \uput[45](-0.5,1.5){$d_i$}
  
  \psline(-1,1.6)(1,1)

  \psline{<->}(0.04,1.4)(0.3,2.4)
	\pscircle[linecolor=black,fillstyle=solid,fillcolor=black](0,1.3){0.1}
   \uput[45](-1.5,0.8){$(x_\bot,y_\bot)$}
	
  \psellipticarc(0,1.22)(0.8,0.8){-11}{75}

 \pscircle[linecolor=black,fillstyle=solid,fillcolor=black](0.34,2.5){0.1}
 
 \pscircle[linecolor=black,fillstyle=solid,fillcolor=black](0,0){0.1}

 \psellipse[linecolor=black,rot=-19.29,linewidth=2pt](0,0)(2.5,1.25)


\psline{->}(-4,-2)(-4,-1)
\psline{->}(-4,-2)(-3,-2)
\uput[45](-3.5,-2.5){$x$}
\uput[45](-4.5,-1.5){$y$}

  \end{pspicture}
    \caption{Orthogonal distance $d_i$ between a measurement $(x_i,y_i)$ and the orthogonal point $(x_\bot,y_\bot)$ on the ellipse.}
    \label{fig:ortho}
\end{figure}
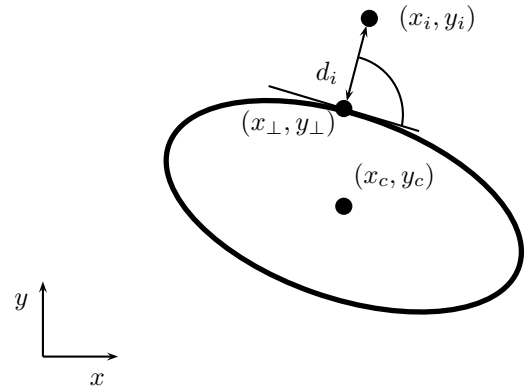

In Section \ref{sec:lsfad} the algebraic distance $C(x,y)$ was minimized. This resulted in a very lean and easy to compute solution for $\mat{p}$.
However, the least-squares solution based on the algebraic distance is not entirely correct.
The error made in this method is called the \textit{high curvature bias}.
Explained in simple terms, the bias arises from the given shape of the ellipse.
Thus, a measurement along the semi-major axis $a$ has a large value for $C(x,y)$ while a measurement along the semi-minor axis $b$ has a smaller value.
However, both measurements can be equally close to the ellipse although their distance $C(x,y)$ is different.
To avoid this bias, the orthogonal distance between a measurement $(x_i,y_i)$ and the ellipse must be used (cf. Figure \ref{fig:ortho}).
The orthogonal distance $d$ is the shortest connection between a point $(x_i,y_i)$ and the ellipse. The parameter vector $\mat{p}$ can be obtained by minimizing the sum of the squared distances
\begin{equation}
\mat{G}(\mat{p})=\sum^{n}_{i=1} d^2_i \enspace .
\end{equation}
For the 2-dimensional case, the distance $d$ is a 4th degree polynomial and can be solved analytically \cite{Chernov2003,ZhengyouZhang.1997}. However, according to \cite{Ahn} the analytic solution is unstable for measurements $|x_i|\approx 0$ or $|y_i|\approx 0$, so in the following section the solution of \cite{Ahn} which uses the Gauss-Newton algorithm will be presented. Once the orthogonal distance $d_i$ for each point $i$ is found, the quality function $\mat{G}$ must be minimized to find the parameter vector $\mat{p}$ with the smallest error. For this purpose the iterative algorithms like: Gauss-Newton, Steepest Gradient Descent, Levenberg-Marquardt or the Simplex algorithm are appropriate.
In Section \ref{subsec:fittingortho}, the Gauss-Newton approach from \cite{Ahn} will be used.

\subsection{Determination of the orthogonal point on the ellipse}
\label{subsec:ortho}
To get the orthogonal distance $d$ the corresponding orthogonal point $(x_\bot,y_\bot)$ on the ellipse is calculated.
The calculation is simplified if this is done in the $XY$ coordinate system, i.e. in the coordinate system in which the ellipse has no rotation and is symmetrical to each coordinate axis.
The elliptic equation for this case is
\begin{eqnarray}
\frac{X^2}{a^2}+\frac{Y^2}{b^2}=1
\end{eqnarray}
or transformed
\begin{eqnarray}
f_1(X,Y) &=&\frac{1}{2} (a^2 Y^2 + b^2 X^2 - a^2 b^2)=0 \enspace .
\end{eqnarray}
For the condition that the tangent of the orthogonal point $(X_\bot,Y_\bot)$ is orthogonal to a measurement $(X_i,Y_i)$
\begin{eqnarray}
\frac{dy}{dx} \cdot \frac{Y_i-Y_\bot}{X_i-X_\bot} = \frac{-b^2 X_\bot}{a^2 Y_\bot} \cdot \frac{Y_i-Y_\bot}{X_i-X_\bot} = -1
\end{eqnarray}
the second equation
\begin{eqnarray}
f_2(X,Y) &=&b^2 X (Y_i-Y) - a^2 Y (X_i -X)=0
\end{eqnarray}
can be derived. These two equations
\begin{eqnarray}
\mat{f} (X,Y) &=& \begin{bmatrix}
       f_1(X,Y)\\[0.3em]
       f_2(X,Y)
     \end{bmatrix} 
     \end{eqnarray}
have to be solved in vector notation. 
Although this is analytically possible, it exhibits numerical instability as already described.
Therefore, in the following the Gauss-Newton method shall be applied to the system of equations.
For this the Jacobian matrix
\begin{eqnarray}
\mat{Q} &=& \begin{bmatrix}
       \frac{\partial f_1}{\partial x} & \frac{\partial f_1}{\partial y}\\[0.3em]
       \frac{\partial f_2}{\partial x} & \frac{\partial f_2}{\partial y}
     \end{bmatrix} \\
&=& \begin{bmatrix}
       b^2 x & a^2 y\\[0.3em]
       (a^2-b^2) y + b^2 y_i & (a^2-b^2)x-a^2x_i 
     \end{bmatrix}
\end{eqnarray}
is calculated. The initial value $\mat{X}_0$ can be calculated by averaging the points $\mat{X}_{k1}$ and $\mat{X}_{k2}$.
\begin{eqnarray}
\mat{X}_0 = \frac{1}{2} (\mat{X}_{k1} + \mat{X}_{k2})
\end{eqnarray}
with 
\begin{eqnarray}
\mat{X}_{k1}= \frac{\begin{bmatrix}
      X_i\\[0.3em]
      Y_i
     \end{bmatrix}ab}{\sqrt{b^2X_i^2+a^2Y^2_i}}
\end{eqnarray}
and
\begin{eqnarray}
\mat{X}_{k2}= \begin{cases}
   \begin{bmatrix}
      X_i\\[0.3em]
      \text{sign}(Y_i) \frac{b}{a} \sqrt{a^2-X_i^2}
     \end{bmatrix} & \text{if} \ |X_i| < a, \\
   \begin{bmatrix}
     \text{sign}(X_i) a\\[0.3em]
     0
     \end{bmatrix}       & \text{if} \ |X_i| \geq a \enspace .
  \end{cases}
\end{eqnarray}
The orthogonal point $\mat{X}'=[X_\bot,Y_\bot]^T$ is calculated iterating
\begin{eqnarray}
\mat{X}'_{k+1} = \mat{X}'_k - \mat{Q}^{-1}_k \mat{f}(X_k) \enspace .
\end{eqnarray}
Finally, by transforming back into the $xy$ coordinate system, the sought vector of the orthogonal distance can be determined
\begin{eqnarray}
\mat{d}_i = \mat{x}_i - \mat{x}'_i \enspace .
\end{eqnarray}
\subsection{Fitting an ellipse with orthogonal distances}
\label{subsec:fittingortho}
With the orthogonal point $\mat{X}'$ from section \ref{subsec:ortho} it is now possible to perform the fitting of the ellipse also with the Gauss-Newton algorithm. For this purpose the vector $\mat{q}=[x_c,y_c,a,b,\alpha]^T$ is defined which contains the searched parameters. 
For each measurement $\mat{x}_i$ a Jacobi matrix
\begin{eqnarray}
\mat{J}_{\mat{x}=\mat{x}'_i} = (\mat{R}^{-1} \mat{Q}^{-1} \mat{B})
\end{eqnarray}
gets computed with
\begin{eqnarray}
\mat{B} &=& \begin{bmatrix} \ \mat{B}_1 \ \mat{B}_2 \ \mat{B}_3 \ \mat{B}_4\end{bmatrix} \enspace , \\
\mat{B}_1&=&\begin{bmatrix}b^2 x C-a^2 y S\\[0.3em]b^2 (yi-y) C+a^2 (xi-x) S \end{bmatrix} \enspace , \\
\mat{B}_2&=&\begin{bmatrix}b^2 x S+a^2 y C\\[0.3em]b^2 (yi-y) S-a^2 (xi-x) C \end{bmatrix} \enspace ,\\
\mat{B}_3&=&\begin{bmatrix}a (b^2-y^2)\\[0.3em]2 a y (xi-x) \end{bmatrix} \enspace ,\\
\mat{B}_4&=&\begin{bmatrix}b (a^2-x^2)\\[0.3em]-2 b x (yi-y) \end{bmatrix}\enspace .\\
\end{eqnarray}
The sought parameter vector $\mat{q}$ is then iteratively determined with a step size $\lambda$ using 
\begin{eqnarray}
\mat{q}_{k+1} = \mat{q}_{k} + \lambda \Delta \mat{q}
\end{eqnarray}
and
\begin{eqnarray}
\begin{bmatrix} \mat{J}_{\mat{x}=\mat{x}'_1} \\[0.3em] \mat{J}_{\mat{x}=\mat{x}'_2} \\[0.3em] \vdots \\[0.3em] \mat{J}_{\mat{x}=\mat{x}'_n}  \end{bmatrix} \Delta \mat{q} = \begin{bmatrix} \mat{d}_1 \\[0.3em] \mat{d}_2 \\[0.3em] \vdots \\[0.3em]  \mat{d}_n\end{bmatrix} .
\end{eqnarray}
\section{Gradient weighted least-squares fitting}
\label{sec:gwls}
For the fitting of ellipses the least-squares solution based on the algebraic distance from Section \ref{sec:lsfad} provides satisfying results.
If accuracy is required, the least-squares solution based on the orthogonal distance from Section \ref{subsec:ortho} should be chosen. 
However, the high accuracy is bought with a considerably higher computational effort. Thus, for each measurement, the orthogonal distance must first be found before iteratively determining the solution for $\mat{q}$. 
In this section, the Gradient Weighted-Least-Squares method will be presented, which is much faster than least-squares based on orthogonal distance and yet more accurate than the least-squares method based on algebraic distance.
The Gradient Weighted-Least-Squares uses a normalization with the help of the gradient which leads to a so-called \textit{statistically optimal weighted algebraic fit} \cite{Chernov2003}.
It is sufficient to divide the original function with its gradient.
Thus, similar to Section \ref{sec:lsfad}, the quadratic goodness measure
\begin{eqnarray}
\mat{G}(\mat{p}) &=& \sum C^{'2}_i (x_i,y_i) \\
&=& \sum \frac{C^2_i(x_i,y_i)}{||\nabla C_i(x_i,y_i)||^2} \\
&=& (\mat{\Phi} \mat{p} -\mat{y})^T \mat{W}^{-1} (\mat{\Phi} \mat{p} -\mat{y})
\end{eqnarray}
can be minimized. Where the normalization is done in the weighting matrix $\mat{W}$, which is 
\begin{equation}
\mat{W}=\text{diag}(||\nabla C_1||^2,||\nabla C_2||^2,...,||\nabla C_n||^2)  \enspace .
\end{equation}
The gradient is calculated from
\begin{equation}
||\nabla C_i(x_i,y_i)||^2 = \bigg(\frac{\partial C_i}{\partial x_i}\bigg)^2+\bigg(\frac{\partial C_i}{\partial y_i}\bigg)^2
\end{equation}
with
\begin{eqnarray}
\frac{\partial C_i}{\partial x_i} &=& 2 \mat{a} x_i + 2 \mat{b} y_i + 2 \mat{d} \enspace , \\
\frac{\partial C_i}{\partial y_i} &=& -2 \mat{a} y_i + 2 \mat{b} x_i + 2 \mat{e} + 2 \mat{y} \enspace .
\end{eqnarray}
This is known as the weighted least-squares method which is calculated by zeroing the first derivative to
\begin{align} 
0  &= \frac{\partial}{\partial \mat{p}} \mat{G}(\mat{p}) \\
0  &= \frac{\partial}{\partial \mat{p}}(\mat{\Phi} \mat{p} -\mat{y})^T \mat{W}^{-1} (\mat{X} \mat{p} -\mat{y}) \\
0 &= 2 \mat{\Phi}^T  \mat{W}^{-1} (\mat{\Phi} \mat{p} - \mat{y})   \\
\mat{p} &= (\mat{\Phi}^T  \mat{W}^{-1} \mat{\Phi})^{-1} \mat{\Phi}^T  \mat{W}^{-1} \mat{y}  \enspace .
\label{eq:gwls}
\end{align}
Thus the solution for $\mat{p}$ looks at first sight like a closed solution for $\mat{p}$. Unfortunately, this is not the case since $\mat{W}$ depends on $\mat{p}$. Therefore, an iterative procedure must be used here, which is structured as follows: 
\begin{enumerate}
\item Calculate an initial value for $\mat{p}^{(k=0)}$ (for example, using the least-squares solution based on the algebraic distance).
\item Calculate the weighting matrix $\mat{W}$ based on the current value of $\mat{p}^{(k)}$ and all measurements.
\item Compute a new $\mat{p}^{(k+1)}$ using the equation \eqref{eq:gwls}
\item If $\mat{p}^{(k)}$ and $\mat{p}^{(k+1)}$ are very similar, then finish the procedure. If not, then continue with step 2.
\end{enumerate}

\section{Robust Estimation}
\label{sec:robust}
In the previous sections, methods were presented which allow the fitting of ellipses. In practice, however, all the methods mentioned have in common that they react very sensitively to outliers in the measurement data. Outliers are incorrect data points or measurements that occur due to various causes. In image processing systems, for example, these are optical reflections that lead to outliers in the feature space. Since these outliers occur unexpectedly in practice, they cannot be modeled exactly or excluded in the conceptual design of a system. Therefore, there is no general approach, but it must be considered from application to application how the estimator can be stabilized against outliers. A number of techniques exist in the literature for this purpose, for example:
\begin{itemize}
\item \textbf{Clustering with the Hough transform}: Here, the measurements are transformed into Hough space and then a maximum likelihood estimator is used to determine the maximum of the cluster. Since it is rarely applicable to up to more than 3 unknowns, it is therefore not suitable for fitting an ellipse.
\item \textbf{Regression Diagnostic}: Here, an initial estimate is made using an arbitrary procedure. Then, individual measurements are removed based on the residuals if they are too large with respect to a threshold. 
\end{itemize}
In the following sections, the M-Estimator and the Least Median of Squares will be considered, since these methods are suitable for fitting ellipses.
\subsection{M-Estimator}
The method of least-squares tries to minimize the squared error. If the measured data have outliers, these are weighted very strongly by the quadratic term. A single outlier can distort the result of an estimator so much that it takes many correct measurements to compensate for this error. M-estimators therefore use an error function $\rho$ that does not grow as much as the quadratic error criterion. However, this property should be the only change to the error function. The error function should still be symmetric, positive definite, and have a minimum at zero. These requirements are met by many functions and it is ultimately a design parameter that can be chosen for any application (a review of these functions can be found in \cite{ZhengyouZhang.1997}). In general, instead of the sum of squared distances $d$, a function $\rho$ is minimized as a function of the distance $d$, formally expressed this is
\begin{equation}
\label{eq:rho}
\text{min} \sum_i \rho (d_i) \enspace .
\end{equation}
However, this quality function is not solved directly but via the so-called \textit{Reweighted Least-Squares} approach. In this approach a vector $\mat{p}$ with $\mat{p}=[p_1,p_2,...,p_n]^T$ is searched to solve the problem of \eqref{eq:rho}. The solution is obtained by setting the first derivative to zero. Furthermore, an \textit{Influence Function} with $\psi(x)=d\rho(x)/dx$ and an \textit{Weight Function} with $w(x)=\psi(x)/x$ are defined. The approach thus has $m$ equations of the form
\begin{eqnarray}
0 &=& \sum^m_i \psi(d_i) \frac{\partial d_i}{\partial p_j} \\
 &=& \sum^m_i w(d_i) d_i \frac{\partial d_i}{\partial p_j} \enspace ,
\end{eqnarray}
that solves the following \textit{Iterated reweighted least-squares} problem 
\begin{eqnarray}
\text{min} \sum^m_i w (d^{k-1}_i) d^2_i .
\end{eqnarray}
This is done by recalculating the weight $w (d^{k-1}_i)$ in each iteration step $k$ based on the previous distance $d^{k-1}_i$ of step $k-1$. 
For many functions for $\psi(x)$ and thus for $w(x)$ no closed-form solution exists and therefore the use of an iterative algorithm is also necessary here (e.g. Newton's method). Furthermore, in most cases the weighted least-squares approach can be used by performing a reweighting in each step.
\begin{table*}[t]
  \centering
    \caption{The parameters of the estimated ellipses compared to the ideal ellipse for the simulation without outliers.}
  \begin{tabular}{l c c c c}
    \hline
    method & semi-major axis $a$ & semi-minor axis $b$ & center $(x_c,y_c)$ & rotation $\alpha$ \\
    \hline
    Ground truth ellipse & 24.00 & 12.00 & $(0.00,0.00)$ & 0.00 \\
    Least-squares fitting using algebraic distance & 21.98 & 11.80 & $(-0.91,0.02)$ & 3.14 \\
    Least-squares fitting using orthogonal distance  & 23.33 & 11.94 & $(-0.53,-0.53)$ & 3.14 \\
		Gradient weighted least-squares fitting  & 22.45 & 11.78 & $(-0.71,0.00)$ & 3.14 \\
    \hline
  \end{tabular}
	\label{table:inliner}

\end{table*}
\subsection{Least Median of Squares}
Another way to handle outliers is to use the median and minimize it:
\begin{eqnarray}
\text{min} \ \text{Median} \ d^2_i \enspace .
\end{eqnarray}
The median is known to be particularly robust against outliers. The problem is that it is very difficult to formulate an analytical formula for the median, since the values have to be ordered by size in order to use the median. Since this is relatively time-consuming for a large data set, only a randomly selected subset of the measurements is used for the estimation. Such a procedure can look as follows:
\begin{enumerate}
\item Choose $m$ random subsets $J_i=\left\{ (x_j,y_j) \right\}^p_{j=1}$ of measurements. Each subset $J_i$ should contain at least $p=5$ measurements, since 5 measurements can be used to determine an ellipse.
	\item For each subset $J_i$, an ellipse $C_i$ is determined using any method. The ellipse $C_i$ is described by a parameter vector $\mat{p}_i$.
	\item For each ellipse $C_i$, the median $M_i$ of the residuals $r$ to each measurement of the entire data set of size $n$ is determined. Formally, this is $M_i=\underset{j=1,...,n}{\text{median}} (r^2_i(\mat{p}_i,\mat{x}_j))$. A choice can be made whether to use the algebraic distance or the more computationally expensive orthogonal distance.
	\item The ellipse $C_i$ and the parameter vector $\mat{p}_i$ with the minimum $M_i$ are chosen as the best estimate.
\end{enumerate}
This procedure is similar to the procedure used by the RANSAC algorithm. It should be noted that the RANSAC algorithm is very popular and many implementations are available. Therefore, for practical use, RANSAC can be preferred to the Least Median of Squares.
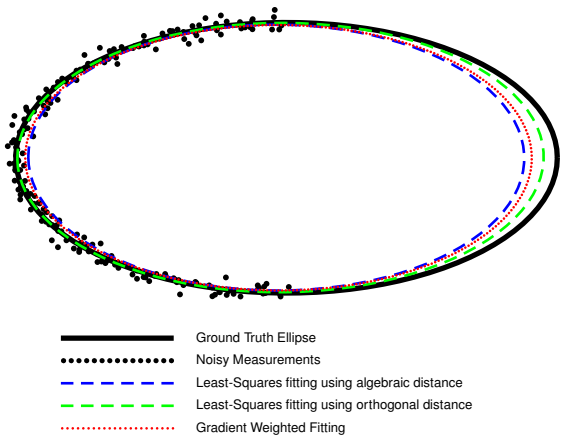
\begin{figure}[htbp]
    \centering
		\psset{xunit=0.15cm,yunit=0.15cm,mathLabel=false}
\begin{pspicture}(-25,-25)(25,15)
 
	\readdata{\gt}{gt.csv}
	\readdata{\noise}{noise.csv}
	\readdata{\LSA}{LSA.csv}
	\readdata{\LSO}{LSO.csv}
	\readdata{\LSWA}{LSWA.csv}
		
\listplot[plotstyle=line,linecolor=black,linewidth=2pt,linestyle=solid,dotsep=0.4pt,yStart=-25,yEnd=25]
						{\gt}

\listplot[plotstyle=dots,linecolor=black,linewidth=0.1pt,linestyle=dotted,dotsep=0.4pt,yStart=-25,yEnd=25,opacity=1]
						{\noise}
		
\listplot[plotstyle=line,linecolor=blue,linewidth=1pt,linestyle=dashed,dotsep=0.4pt,yStart=-25,yEnd=25,opacity=1]
						{\LSA}
						
\listplot[plotstyle=line,linecolor=green,linewidth=1pt,linestyle=dashed,dotsep=0.4pt,yStart=-25,yEnd=25,opacity=1]
						{\LSO}
						
\listplot[plotstyle=line,linecolor=red,linewidth=1pt,linestyle=dotted,dotsep=0.4pt,yStart=-25,yEnd=25,opacity=1]
						{\LSWA}

		
		
		\rput(-25,-25){
			
			\psline[plotstyle=line,linecolor=red,linewidth=1pt,linestyle=dotted,dotsep=1pt,yStart=-25,yEnd=25,opacity=1](5,1)(15,1)
			\rput[l](17,1){\sffamily \tiny Gradient Weighted Fitting}

			\psline[plotstyle=line,linecolor=green,linewidth=1pt,linestyle=dashed,dotsep=1pt,yStart=-25,yEnd=25,opacity=1](5,3)(15,3)
			\rput[l](17,3){\sffamily \tiny Least-Squares fitting using orthogonal distance}
			
			\psline[plotstyle=line,linecolor=blue,linewidth=1pt,linestyle=dashed,dotsep=1pt,yStart=-25,yEnd=25,opacity=1](5,5)(15,5)
			\rput[l](17,5){\sffamily \tiny Least-Squares fitting using algebraic distance}
			
			\psline[plotstyle=dots,linecolor=black,linewidth=2pt,linestyle=dotted,dotsep=1pt,yStart=-25,yEnd=25,opacity=1](5,7)(15,7)
			\rput[l](17,7){\sffamily \tiny Noisy Measurements}
			
			\psline[linecolor=black,linewidth=2pt,linestyle=solid,dotsep=0.4pt](5,9)(15,9)
			\rput[l](17,9){\sffamily \tiny Ground Truth Ellipse}
		}
  \end{pspicture}
    \caption{Qualitative result of the simulation. It can be clearly seen that the least-squares method based on the orthogonal distance estimates the ellipse best. The least-squares estimation based on the algebraic distance has the highest error.}
    \label{fig:inliner}
\end{figure}
\section{Simulation}
\label{sec:simulation}
This section aims to demonstrate the performance of the algorithms shown.
The simulation code is publicly available at \textit{github.com/sebdi/ellipse-fitting}, which allows any reader to reproduce the results or run simulations with different parameters.

\subsection{Fitting with noisy measurements without outliers}
Noisy measurements were generated using equation \eqref{eq:paraForm1} and \eqref{eq:paraForm2}. 
For this, the parameter $t$ is varied in the interval $[\frac{\pi}{2},\frac{3 \pi}{2}]$ and homogeneous-isotropic noise with constant variance $\sigma^2=0.25$ is added to $x(t)$ and $y(t)$.
Then, an estimate of the ellipse is made using these measurements. 
For this the presented methods
\begin{itemize}
\item least-squares with algebraic distance,
\item least-squares with orthogonal distance,
\item and gradient weighted least-squares.
\end{itemize} is used. 
Figure \ref{fig:inliner} shows the qualitative result of such a simulation.
Table \ref{table:inliner} further shows the parameters of the ellipse and the estimates of each method.
The simulation shows that the best result is obtained with the least-squares method based on the orthogonal distance.
The estimate based on the algebraic distance has the largest error, followed by the Gradient Weighted Least-Squares method which is a good compromise between the other two methods. 

\subsection{Fitting with noisy measurements with outlier}
\begin{figure}[htbp]
\vspace{10mm}
    \centering
		\psset{xunit=0.15cm,yunit=0.15cm,mathLabel=false}
\begin{pspicture}(-25,-29)(35,13)
 
	\readdata{\gt}{out_gt.csv}
	\readdata{\noise}{out_noise.csv}
	\readdata{\LSA}{out_LSA.csv}
	\readdata{\LSO}{out_LSO.csv}
	\readdata{\LSWA}{out_GWLS.csv}
	\readdata{\MEst}{out_MEst.csv}	
	\readdata{\LSoS}{out_LSoS.csv}		
\listplot[plotstyle=line,linecolor=black,linewidth=2pt,linestyle=solid,dotsep=0.4pt,yStart=-25,yEnd=25]{\gt}

\listplot[plotstyle=dots,linecolor=black,linewidth=0.1pt,linestyle=dotted,dotsep=0.4pt,yStart=-15,yEnd=25,opacity=1]{\noise}
		
\listplot[plotstyle=line,linecolor=red,linewidth=1pt,linestyle=dotted,dotsep=0.4pt,yStart=-25,yEnd=25,opacity=1]{\LSO}
						
\listplot[plotstyle=line,linecolor=red,linewidth=1pt,linestyle=dashed,dotsep=0.4pt,yStart=-25,yEnd=25,opacity=1]{\MEst}
						
\listplot[plotstyle=line,linecolor=blue,linewidth=1pt,linestyle=dashed,dotsep=0.4pt,yStart=-25,yEnd=25,opacity=1]{\LSoS}
						
\listplot[plotstyle=line,linecolor=blue,linewidth=1pt,linestyle=dotted,dotsep=0.4pt,yStart=-25,yEnd=25,opacity=1]{\LSA}

\listplot[plotstyle=line,linecolor=green,linewidth=1pt,linestyle=dotted,dotsep=0.4pt,yStart=-25,yEnd=25,opacity=1]
						{\LSWA}

		
		
		\rput(-25,-25){
			
			\psline[plotstyle=line,linecolor=blue,linewidth=1pt,linestyle=dashed,dotsep=0.4pt,yStart=-25,yEnd=25,opacity=1](5,-3)(15,-3)
			\rput[l](17,-3){\sffamily \tiny Least Median of Squares}

			\psline[plotstyle=line,linecolor=red,linewidth=1pt,linestyle=dashed,dotsep=0.4pt,yStart=-25,yEnd=25,opacity=1](5,-1)(15,-1)
			\rput[l](17,-1){\sffamily \tiny M-Estimator (Cauchy)}

\psline[plotstyle=line,linecolor=green,linewidth=1pt,linestyle=dotted,dotsep=0.4pt,yStart=-25,yEnd=25,opacity=1](5,1)(15,1)
		\rput[l](17,1){\sffamily \tiny Gradient Weighted Least Squares Fitting}		
			
			\psline[plotstyle=line,linecolor=red,linewidth=1pt,linestyle=dotted,dotsep=0.4pt,yStart=-25,yEnd=25,opacity=1](5,3)(15,3)
		\rput[l](17,3){\sffamily \tiny Least-Squares Fitting using orthogonal distance}
			
			\psline[plotstyle=line,linecolor=blue,linewidth=1pt,linestyle=dotted,dotsep=0.4pt,yStart=-25,yEnd=25,opacity=1](5,5)(15,5)
			\rput[l](17,5){\sffamily \tiny Least-Squares Fitting using algebraic distance}
			
		\psline[plotstyle=dots,linecolor=black,linewidth=2pt,linestyle=dotted,dotsep=1pt,yStart=-25,yEnd=25,opacity=1](5,7)(15,7)
			\rput[l](17,7){\sffamily \tiny Noisy Measurements with Outlier}
			
			\psline[linecolor=black,linewidth=2pt,linestyle=solid,dotsep=0.4pt](5,9)(15,9)
			\rput[l](17,9){\sffamily \tiny Ground Truth Ellipse}
		}
  \end{pspicture}
    \caption{Qualitative result of the simulation with measurements that have outliers. It can be clearly seen that methods like M-Estimator and Least Median of Squares achieve a better result because they take into account outliers in the measurement data.}
    \label{fig:outliner}
\end{figure}
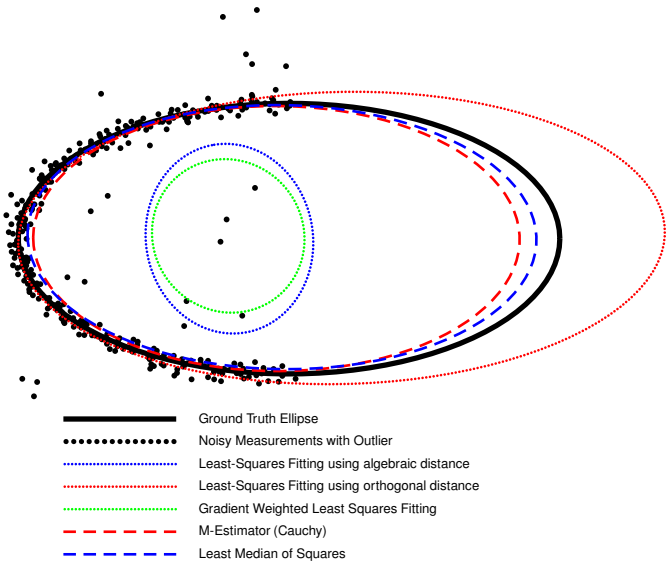

\begin{table*}[t]
  \centering
   \caption{The parameters of the estimated ellipses compared to the ideal ellipse for the simulation with outliers.}
  \begin{tabular} {l c c c c}
    \hline
    method & semi-major axis $a$ & semi-minor axis $b$ & center $(x_c,y_c)$ & rotation $\alpha$ \\
    \hline
    Ground truth ellipse & 24.00 & 12.00 & (0.00,0.00) & 0.00 \\
    Least-squares fitting using algebraic distance& 8.44 & 7.40 & (-5.28,-0.00) & 0.00 \\
		Least-squares fitting using orthogonal distance & 28.67 & 12.96 & (4.65,4,65) & 3.16 \\
		Gradient weighted least-squares fitting  & 6.94 & 6.63 & (-5.39,0.24) & 0.00 \\
		M-Estimator (Cauchy)  & 21.56 & 11.76 & (-1.12,0.01) & 0.00 \\
		Least median of squares & 22.56 & 11.71 & (-0.625,0.13) & 3.13 \\
    \hline
  \end{tabular}
	\label{table:outliner}
 
\end{table*}
As mentioned earlier, measured data often have outliers that are not accounted for in the standard procedures from Section \ref{sec:lsfad} to \ref{sec:gwls}.
In this simulation, some outliers were added to the measurements to highlight the problem.
In addition to the previous methods,
the methods
 \begin{itemize}
 \item M-Estimator
 \item Least Median of Squares
 \end{itemize}
were simulated, that take outliers into account.
The qualitative result is shown in Fig. \ref{fig:outliner} to be examined.
The estimated parameters of the ellipses can be found in Tab. \ref{table:outliner}.
The simulation shows that the methods that take outliers into account are much more robust compared to the standard methods.

\section{CONCLUSION}
\label{sec:zus}
In this work, methods were discussed with which ellipses can be approximated on the basis of noisy measurements. For this purpose, a very simple method was presented with the least-square solution based on the algebraic distance. However, this method is inaccurate because it has the so-called \textit{High Curvature Bias}. This problem is addressed by the orthogonal distance, but its solution is more computationally expensive. Furthermore, the Gradient Weighted-Least-Squares approach was presented as a method that, although it uses the algebraic distance, achieves good accuracy. However, the Gradient Weighted-Least-Squares solution is much faster to compute than the least-squares method based on the orthogonal distance. 

Furthermore, robust methods were presented that are less sensitive to outliers in the measurements. For this purpose, the M-estimators use a different error function that weights outliers less than the squared error measure. The other method is to use the median where large outliers are ignored. Last but not least, the theoretical explanations were demonstrated with the help of a simulation.

All methods shown have in common that they consider the problem of noisy measurements as a deterministic problem. However, there are also stochastic methods such as \cite{Baum,Baum2} that solve the problem using a Bayesian approach. In practice, therefore, it must be weighed whether to use a simple and fast method such as the least-squares approach based on algebraic distance or a more computationally expensive approach. For this purpose, the simulation provided on github.com can be used to make initial attempts to determine which method is more suitable for the application problem.

Also, the methods shown do not have a recursive structure since it is assumed that all measurements are already available. If measurements only occur sequentially, for example in tracking, then recursive methods such as the Kalman filter or, for ellipses, the extended Kalman filter are necessary.

\bibliographystyle{IEEEtran}
\bibliography{Bibliography}
\end{document}